\documentclass[article]{elsarticle}
\usepackage{xcolor}
\usepackage{hyperref}
\usepackage{amsmath}
\usepackage{mathtools}
\usepackage{amssymb}
\usepackage{colortbl}
\usepackage[caption = false]{subfig}

\begin{document}

\begin{frontmatter}

\title{Automatic Identification of MHD Modes in Magnetic Fluctuations Spectrograms using Deep Learning Techniques}

\author[tecno]{A. Bustos}
\ead{andres.bustos@ciemat.es}
\author[lnf]{E. Ascas\'ibar}
\author[lnf]{A. Cappa}
\author[tecno]{R. Mayo-Garc\'ia}

\address[tecno]{Department of Technology - CIEMAT, Av. Complutense 40, 28040 Madrid, Spain}
\address[lnf]{National Fusion Laboratory - CIEMAT, Av. Complutense 40, 28040 Madrid, Spain}

\begin{abstract}
The control and mitigation of MHD oscillations modes is an issue in fusion science because they can contribute to the outward particle/energy flux and can drive the device away from ignition conditions. It is then of general interest to extract the mode information from large experimental databases in a fast and reliable way. We present a software tool based on Deep Learning that can identify these oscillations modes taking Mirnov coil spectrograms as input data.  It uses Convolutional Neural Networks that we trained with manually annotated spectrograms from the TJ-II stellarator database. We have tested several detector architectures, resulting in a detector AUC score of 0.99 on the test set. Finally, it is applied to find MHD modes in our spectrograms to show how this new software tool can be used to mine large databases.
  
\end{abstract}


\end{frontmatter}


\section{Introduction}
\label{sec_Introduction}

The confinement of energetic/fast ions is one of the major issues in fusion science, since those ions, either alphas produced by the fusion reactions or proton/deuterons produced or accelerated by the different heating systems, are the main source of plasma heating and current drive in reactor scenarios. Their dynamic is strongly influenced by the magnetic field because the radial particle drift depends on the product of the ion energy and other functions of the magnetic field \cite{Helander_book, Balescu_book}. Unfortunately, the presence of fast ions in a magnetized plasma can destabilize magneto-hydrodynamic [MHD] oscillations that interact with them and can expel them from the plasma bulk. This leads to an important outward heat flux, either towards the vacuum vessel or any other hardware in the device. Apart from the energy loss that drives the system away from ignition conditions, the device itself can suffer physical damage and the plasma may be polluted with impurities caused by sputtering processes. Among the several types of MHD instabilities, one which is easily destabilized in typical toroidal devices is the well-known shear Alfv\'en wave \cite{Vlad_1999}. It is indeed of great interest to understand both theoretically and experimentally the MHD activity in order to control or mitigate the fast ion losses. Mirnov coils \cite{Mirnov_Soviet_1971} are diagnostics that can measure magnetic field fluctuations associated to the Alfv\'en modes in specific spatial locations with a high sampling rate ($> 1$ MHz).

A quantitative experimental study of these MHD oscillations requires the identification of series of triplets $(t_i, f_i, s_i)$ that define one of those modes, being $i$ an integer index that runs though the mode lifetime. Here, $t_i$ stands for time, $f_i$ is the dominant frequency at $t_i$ and $s_i$ is the mode intensity. This information is enough to characterize the oscillation mode and to relate it to other plasma parameters. We have attempted to extract this data from a few Mirnov coil spectrograms by means of two classical techniques: standard image thresholding and Otsu's thresholding method. Both methods take a grayscale image with pixel intensity in the range from 0 to 255 and produce a binary black and white image. Those white pixels are identified as important oscillations in the spectrogram and the black pixels to the background. In standard thresholding one simply defines a global threshold to the pixel intensity and classifies the pixels in black or white according ot this value. Otsu's thresholding method is more sophisticated \cite{Otsu_1979}. It applies a clustering method to the pixel intensity distribution to divide it in two subsets in such way that the sum of the variances of each subset is minimum.

\begin{figure}[ht]
  \centering
  \includegraphics[width=12cm]{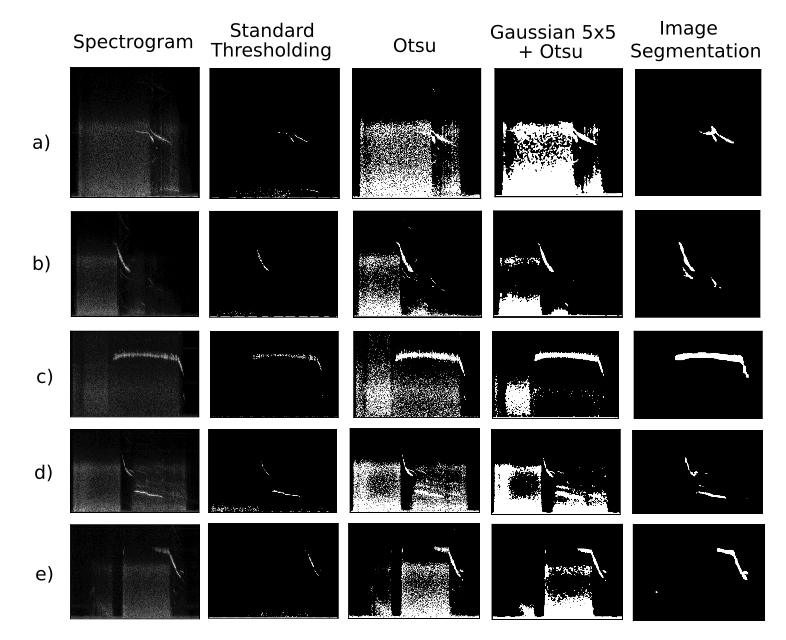}
  \caption{Application of standard thresholding methods to five different Mirnov coil spectrograms.}
  \label{fig_comparison_Otsu}
\end{figure}

In Fig. \ref{fig_comparison_Otsu} we depict the application of these techniques to five spectrograms taken from the TJ-II database up to 500 kHz. Each row of images corresponds to one spectrogram, which is shown on the first column. The second column shows the result of appliying standard thresholding with a threshold value of 127. In the third and forth columns we apply Otsu's method to the original spectrogram and to the spectrogram smoothed by a 5x5 Gaussian filter. Finally, in the last column we show the results of the mode identification obtained with the method presented in this paper (Image Segmentator). Clearly the classical methods produce very noisy results and they identify as modes a large fraction of the background noise. One may think a possible improvement would be to cut the spectrograms in the low frequency range where the noise is more intense, but unfortunately some modes also live in this frequency range (spectrograms d) and e)). The difficulties to establish objective conditions to detect the modes and apply them to a large number of spectrograms motivates us to try to solve this problem using Artificial Intelligence.

Thus, we have developed a software tool based on Deep Learning that automatically identifies modes in a Mirnov coil spectrogram. It is designed to take as input a spectrogram (RGB image, for reasons explained in the next sections) and produce a B/W image with the same resolution, where the white pixels indicate the existence of MHD activity. In the context of supervised learning, we have annotated a dataset of 500 spectrograms and trained a neural network to identify the interesting modes in those images. To illustrate the potential of the detector, we apply it to the whole dataset and build a database of modes. Then we can obtain information of the distribution of physical features of the population of modes, such as their frequency values, duration, intensity, etc. The physics underlying a dataset of spectrograms can now be faster derived, being possible to quickly and automatically identify the modes characteristics and behavior. This is of utmost importance when the analysis has to be made with a big dataset, like the TJ-II stellarator database, that amounts up to more than 50,000 discharges. Moreover, the methodology proposed in this work can be easily adapted to any other dataset of spectrograms from any other fusion device worldwide.

The reminder of this paper is organized as follows: in Section \ref{sec_methodology} we briefly review some basic concepts of computer vision and Deep Learning, explaining the detection methodology and how we created the spectrograms dataset. In Section \ref{sec_training} we describe the way the neural network is trained to perform the mode identifications and in Section \ref{sec_database} we apply it to the whole dataset to illustrate its applications. Finally, in Section \ref{sec_conclusions} we present our conclusions and next steps in this line of research.


\section{Methodology}
\label{sec_methodology}

Deep Learning \cite{Goodfellow_book} comprises modern numerical tools and algorithms based on deep neural networks. It is used in disparate fields to automatize human work. Relevant examples of the use of Deep Learning are autonomous driving, face recognition, recommender systems, anomaly detection, and treatment of medical data among others. In the last years it has proven to be very useful in those tasks and more, and can produce excellent results in problems that are difficult to code or express mathematically. This is clearly the present case because, as we mentioned in Sec. \ref{sec_Introduction}, it is very difficult to define objective criteria of what regions of the spectrogram correspond to oscillation modes.

Deep Learning uses artificial neural networks with diverse architectures and depths depending on the dataset and task assigned. There are usually several millions of parameters to adjust in the neural network. In the paradigm of supervised learning, the neural network uses a dataset of input and labeled output events to minimize a cost function and then predict the output of unlabeled data. The fit of the neural network parameters is known as the \emph{learning phase}, an iterative procedure with high computational cost. The proliferation of Graphical Processing Units [GPU's] for scientific computation has shortened drastically the training time for most Deep Learning problems, spreading its use and pushing the R\&D in new architectures and algorithms.

Starting from a Mirnov coil spectrogram, the mode detection problem on the image can be treated as a pixel classification problem, where the probability of being part of a mode is assigned to each pixel of the spectrogram. This is compatible with an image segmentator architecture.

\subsection{Image Segmentator}

An image segmentator is a kind of neural network with an encoder-decoder architecture that classifies the pixels of an input image in a predefined set of classes. The encoder part takes the image and, though a set of convolutional and pooling layers, reduces the image to a latent representation in a lower dimension space. Recall that an RGB image has $3\times N_x \times N_y$ dimensions, being $N_x$ and $N_y$ the image resolution in pixels. The decoder part of the segmentator takes the latent representation and creates an output image with the same resolution using multiple deconvolutional and dense layers. Finally, the segmentator produces a series of $N_x \times N_y$ arrays equal to the number of classes. Each matrix element represents the probability that a pixel corresponds to a class in the input image. This last part of the segmentator depends on the number of classes we want to classify. For this work, a binary classificator is used, since we just have two pixel classes: mode and no-mode. A comprehensive review of these techniques and neural networks can be found in \cite{Minaee2020ImageSU}.

As we mentioned before, image segmentators have many numerical parameters (weights), in the order of $10^7-10^8$. If we start from a random initialization, we would require a huge annotated dataset of images and great computing capabilities. Fortunately, it is possible to obtain the pre-trained weights of several encoder architectures, saving time and resources. We have tried three different encoders that are well established and known in the field of computer vision, namely VGG \cite{Simonyan2015VeryDC}, MobileNet \cite{howard2017mobilenets} and Vanilla (alias for simple, direct CNN encoder without fancy features).VGG and MobileNet are pre-trained with large image databases of tens of thousands of images, whose weights can be downloaded from public repositories. On the other hand, the simpler Vanilla encoder has to be trained from random weights initialization. The decoder part, responsible for the pixel classification, must be trained with our spectrogram dataset. We have used two different decoders: FCN8 and FCN32. The weights of the decoder part are more task-dependent so they are randomly initialized. The number of free parameters to fit the neural network depends mostly on the encoder, being around 70, 124, and 226 million parameters for Vanilla, VGG and MobileNet respectively.

An implementation (in Python) of those networks can be found in \cite{repo_segmentator} and it was adapted to our purposes. All computations have been done using NVIDIA Tesla P100 GPUs with 3,584 CUDA cores and 12 GB of memory located in the ACME cluster at CIEMAT. The library versions are Tensorflow-gpu 1.15, CUDA 10.0 and cuDNN 7.

\subsection{Dataset Gathering}

We have applied this mode detection method to the magnetic fluctuations present in the TJ-II stellarator \cite{Jimenez-Gomez_FST_2007}. TJ-II is a medium size flexible heliac with major radius equal to 1.5 m and minor radius of the order of 0.25 m, depending on the magnetic configuration. TJ-II has several diagnostics that measure plasma density, temperature, voltage, energy contained, magnetic fluctuations, impurity concentration \dots in its database of $\sim 50,000$ discharges.

One of the TJ-II Mirnov coils, labeled MIR5C, has been extensively used in the present study due to its particularly good signal-to-noise ratio. It is located inside the vacuum vessel, around 4 cm away from the last close flux surface of the plasma and protected by a thin-walled stainless steel tube. It is positioned in a $\varphi$ = 298 degrees toroidal plane and it is oriented so as to measure primarily the poloidal component of the magnetic field created by the plasma. The corresponding measured voltage is digitized with 1 MHz sampling rate \cite{Jimenez-Gomez_NF_2011}. 

We firstly choose a dataset of $\sim 500$ TJ-II discharges to calculate the Mirnov coil spectrograms, annotate and train our mode detector. We have made sure that it contains images of a variety of TJ-II scenarios: discharges with known Alfv\'en activity, discharges with no activity, discharges with different heating methods and failed discharges. Failed discharges are discharges that did not lead to a viable plasma for experimenting because of any reason. Thus we made our dataset representative of the TJ-II database in terms of magnetic fluctuations. The time interval used to calculate the spectrogram from the Mirnov coil signal is 1 ms, more than two orders of magnitude shorter than the typical TJ-II discharge and also shorter than the typical mode frequency evolution times. With this we counted on 256 Fourier modes reaching up to 500 kHz, high enough to capture the typical MHD oscillation frequencies.

We then had to annotate, i.e. create the ground truth of all the 500 spectrograms to identify the modes activity and train the neural network. The annotation process was done manually using any simple picture editor, freely available in all operative systems. The intention was that the image segmentator would mark and identify regions of the spectrograms in a similar way as the humans do. We have annotated the images in a rough way, marking approximately the surface of the image that contains a mode that we considered interesting. This is a subjective task and probably two people would not annotate an image in the same way. We intentionally tended to ignore MHD activity in the low frequency region of the spectrogram because we focused on the identification of shear Alfv\'en waves which usually appear above 50 kHz for the typical plasma density, current and magnetic geometry of the TJ-II stellarator.

All modes/structures that appeared in each spectrogram were annotated, not only the Alfv\'en modes because it is not possible to identify them only with the spectrogram information. The Alfv\'en modes will be identified later in a post processing stage. We must mention that the annotation has been done by all the authors, in order to introduce diversity and reduce human biases as much as possible. This indeed would make the mode detector more flexible and robust.

\begin{figure}[ht]
  \centering
  \includegraphics[width=8cm]{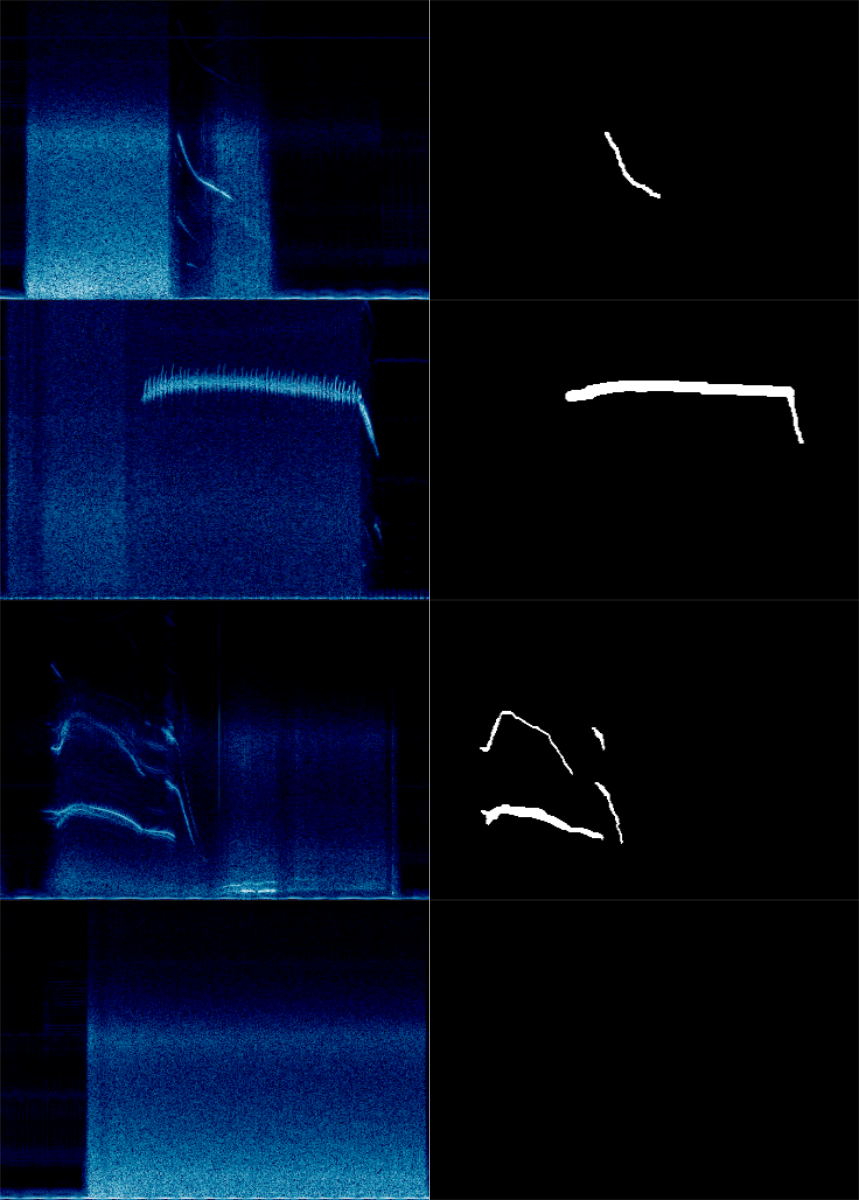}
  \caption{Four examples of manually annotated spectrograms. Notice the lack of annotations in the last picture because no mode is visible on the spectrogram.}
  \label{fig_examples_annotation}
\end{figure}

Finally we ended up with a dataset of 500 pairs of images. Each pair contains an RGB spectrogram of the Mirnov coil and a B/W image that represents the ground truth. In Fig. \ref{fig_examples_annotation} we can see four examples of the dataset.

\subsection{Performance Measurements}
In order to compare between several segmentator models and to give an objective performance indicator we measure the performance of our segmentator with two different quantities: the Intersection over Union [IoU] and the Area Under the receiver operating characteristic Curve [AUC]. Here we must emphasize that our dataset is unbalanced in terms of the classification classes, and that the performance metrics must consider this. Some other indicators, like the standard categorical accuracy, may lead to confusing conclusions.

The IoU, also known as Jaccard Index, is a measure of similarity between two images and takes values in $[0,1]$.  The two images compared are in our case the ground truth and the prediction of an image in the test set. IoU is defined as the area of overlap divided by the area of union of the regions of the image classified as modes. Then we can quantify the goodness of the mode detector with the average of all IoU's of all events in the test set.

The AUC is a more sophisticated measure \cite{BRADLEY19971145}, also in the  $[0,1]$ range. The receiver operating characteristic curve is created by plotting the true positive rate [TPR] versus the false positive rate [FPR] at various detection threshold settings. Here we understand as positive a pixel that belongs to a mode and negative otherwise). This threshold is the minimum probability that the segmentator must give to classify this pixel as positive. It is usually fixed at 0.5, but if we vary this number between 0 and 1 we obtain a curve of (TPR, FPR) points. The Area Under the Curve [AUC] is shown to be a good measure of the segmentator quality. Values of AUC $\sim 0.95$ are generally considered good enough, while AUC $\sim 0.99$ would be excellent.

\section{Training the Model}
\label{sec_training}
The standard procedure in machine learning projects is to divide the dataset in three subsets (train, validation and test sets) in order to separate the learning and evaluation processes with objectivity and statistical independence. The whole dataset is divided randomly so the subsets have the same distribution, and their relative size is taken to be 70/15/15 for train, validation and test.

The training set is used to iteratively adjust the weights of the image segmentator so the loss function is minimized. We chose the categorical cross-entropy of a binary classificator as the loss function. All along the learning phase, the final model is the one model that minimizes the loss function on the validation set. Then, we used the test set to evaluate the model using the  aforementioned performance metrics.

We trained the six combinations of encoders and decoders and compare them in terms of the metrics IoU and AUC. We used an ADAM optimizer \cite{kingma2014method} to avoid getting stuck in local minima of the loss function. Around 200 different trainings are performed, varying not only the network architecture but also other numerical parameters as the learning rate [$lr$], batch size [$bs$] and the number of epochs. The learning rate is a parameter that controls the amount that the weights are updated during the training phase.The batch size is the number of images that are backpropagated simultaneously through the network to smooth the loss function gradiant. One epoch denotes a complete pass of the training set.

As an example of the convergence studies executed, in Table \ref{table_comparative_models} we compare the six architectures fixing $lr=0.0005, \, bs=8$ and training for 500 epochs. The best model corresponds to a VGG encoder with FCN8 decoder. This is indeed the best model found in all convergence studies, so it is used in the reminder of this paper.  We also show some additional information: the training time and the categorical accuracy of the detector.

\begin{table}
  \begin{centering}
    \begin{tabular}{ |c|c|c|c|c|c| }
      \hline
      \rowcolor[gray]{0.9} Encoder & Decoder & Tr. time [min] & IoU & AUC & Cat. Acc.\\
      \hline
      Vanilla   & FCN32  & 100 & 0.233 & 0.976 & 0.989\\
      VGG       & FCN32  & 193 & 0.263 & 0.979 & 0.989\\
      MobileNet & FCN32  & 391 & 0.260 & 0.973 & 0.988\\
      Vanilla   & FCN8   & 99  & 0.401 & 0.987 & 0.991\\
      VGG       & FCN8   & 187 & 0.427 & 0.991 & 0.991\\
      MobileNet & FCN8   & 388 & 0.392 & 0.988 & 0.990\\
      \hline
    \end{tabular}
    \caption{\label{table_comparative_models} Training time and performance metrics for several image segmentator architectures ($lr=0.0005,\,bs=8,$ num. epochs=500). We can see here that the Categorical Accuracy is not the best performance indicator in this unbalanced dataset because it does not vary much and it is very close to 1 in comparison with the AUC and specially IoU.}
  \end{centering}
\end{table}

\begin{figure}[ht]
  \centering
  \includegraphics[width=12cm]{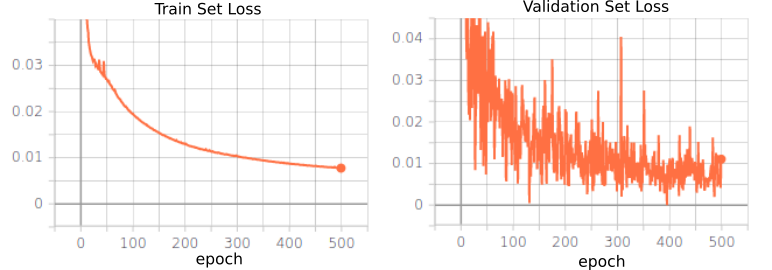}
  \caption{Learning curves (loss function VS epoch) of the training and validation sets.}
  \label{fig_learning_curves}
\end{figure}

Fig \ref{fig_learning_curves} shows the learning curves of the loss functions in the training and validation sets as a function of the epoch during the learning stage. We can see that the 500 epochs are almost enough to stabilize and that the loss in the validation set does not increase as the train loss decreases. This means that the training time is long enough and that no overfitting is present. Training for longer time (2,000 epochs) does not give significant improvement.

\begin{figure}[ht]
  \centering
  \includegraphics[width=8cm]{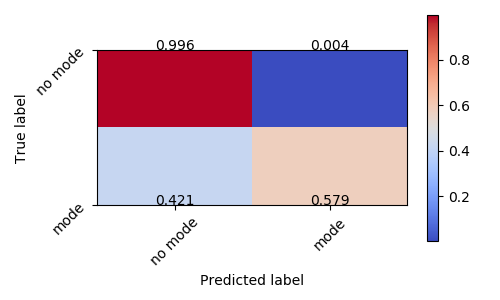}
  \caption{Normalized confusion matrix of the mode detector at the pixel level.}
  \label{fig_confusion_matrix}
\end{figure}

\begin{figure}[ht]
  \centering
  \includegraphics[width=8cm]{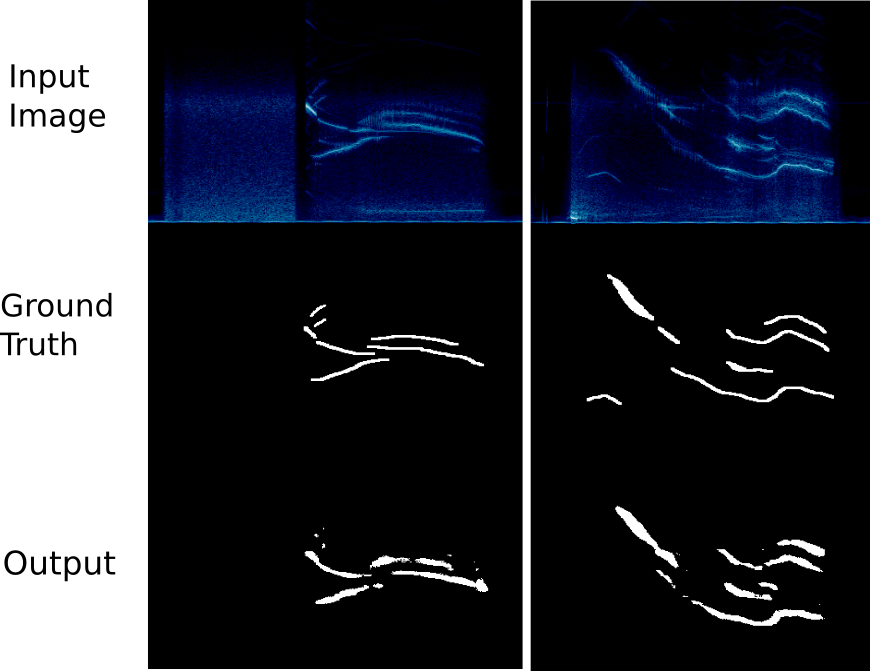}
  \caption{Input, ground truth and raw output of the mode detector for shots \#32912 (left) and \#38940 (right) in the test set. Noisy data appears in some regions and it must be cleaned.}
  \label{fig_output_raw}
\end{figure}

The normalized confusion matrix at the pixel level is plotted in Fig. \ref{fig_confusion_matrix}. It provides more detailed information on the classification of the pixels. Although an IoU = 0.427 and a True Positive Ratio = 0.579 may look small, we will see in the next section that, with some data post processing, we have built an effective mode detector. The reason for this is that the mode detector produces noisy data, which must be smoothed and filtered before identifying a region of the spectrogram as a mode. In Fig. \ref{fig_output_raw} we can see some examples of these noisy images.

\section{Massive Analysis of Spectrograms and Database Creation}
\label{sec_database}

In this section we describe the use of all the 500 spectrograms of the dataset as input images on the trained segmentator, the post processing of its output and the application of descriptive statistics to characterize the oscillations modes and locate the Alfv\'en activity. This post processing stage, including the feed forward of the neural network on the complete dataset, takes around 60 minutes on a common desktop computer with no GPU. 

\subsection{Post processing}
We first apply a blob detector to the segmentator output image (implemented in the opencv library). This identifies connected regions of the spectrogram with activity.In order to remove noise and find relevant modes, we disregarded all blobs shorter than certain minimum duration, say 10 ms. Then, for each mode/blob we calculated a list of time-frequency $(t_i, f_i)$ pairs that represents it. Fixing a time interval in $t_i$, we calculated $f_i$ as the average frequency weighted with the spectrogram values. This makes the frequency estimator very robust with the possible fluctuations of the upper/lower borders of the blob. 

At this point we had a numerical description of every individual mode as a frequency time series. This is the first time that this information was extracted from the TJ-II database and can be used for multiple purposes. For example, we can study the Alfvenic character of all modes in the database using the correlation coefficient of the mode frequency with $n^{-1/2}$.

\subsection{Dataset analysis}
\begin{figure}[ht]
  \centering  \includegraphics[width=12cm]{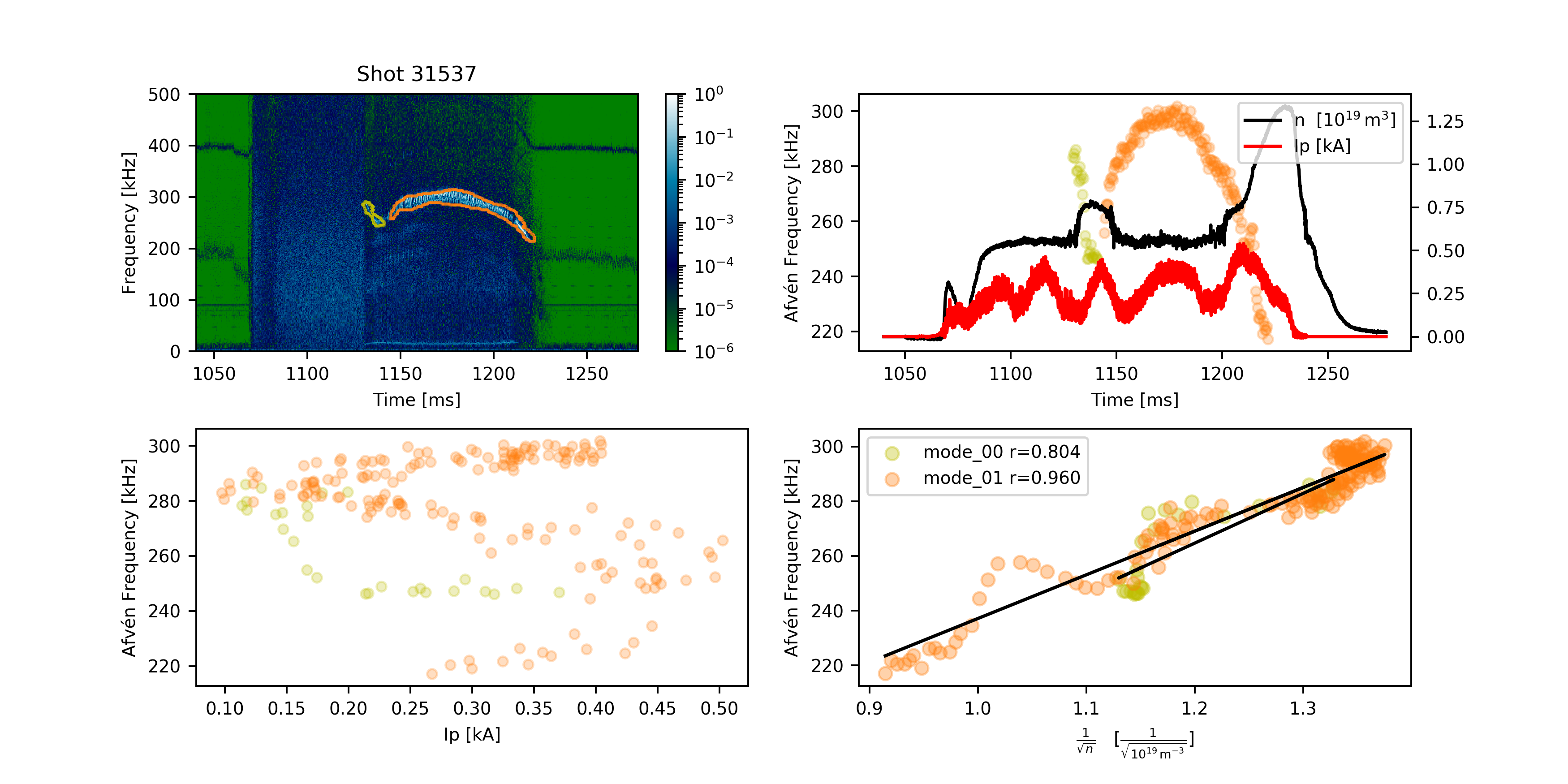}
  \includegraphics[width=12cm]{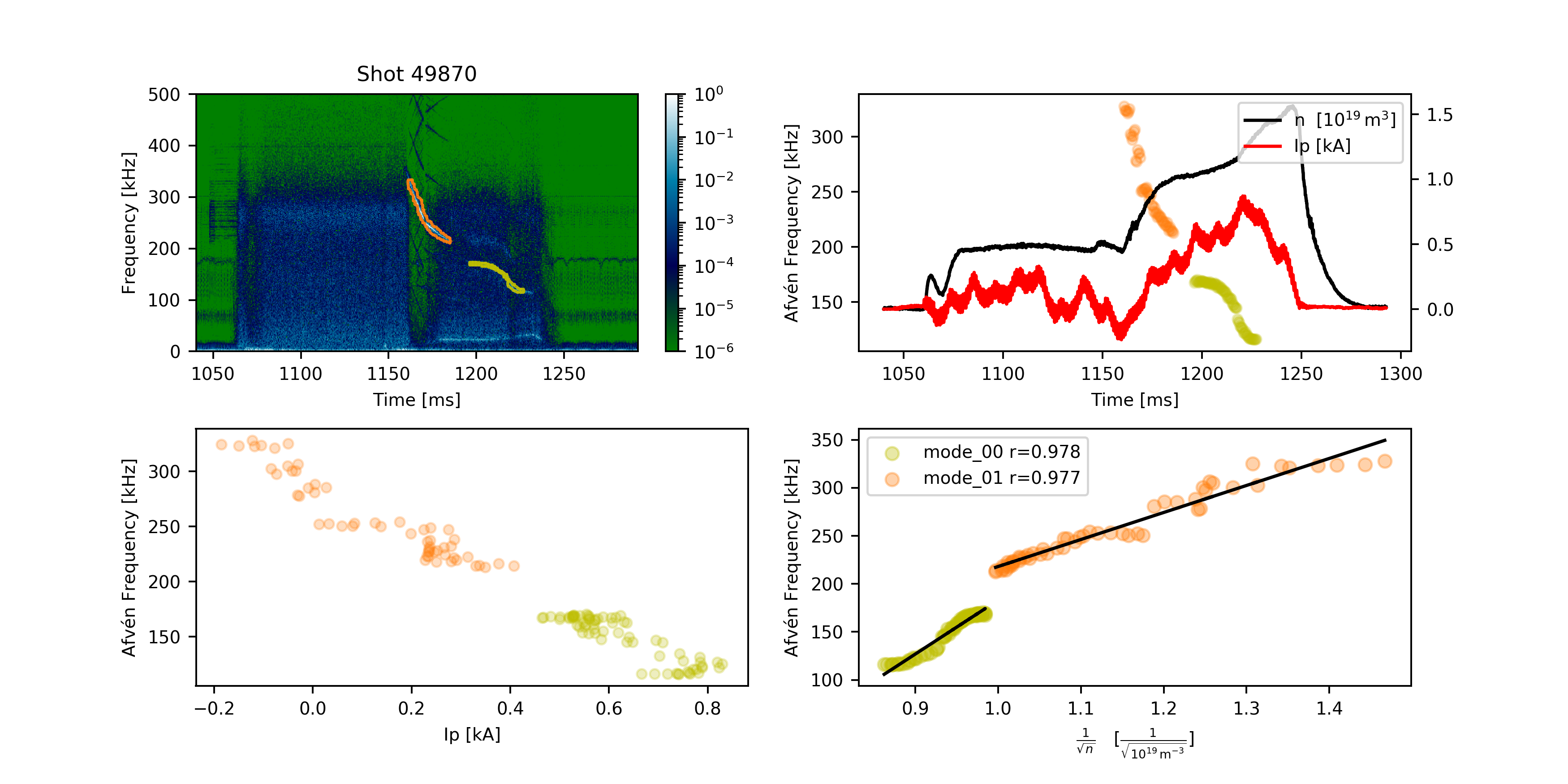}
    \includegraphics[width=12cm]{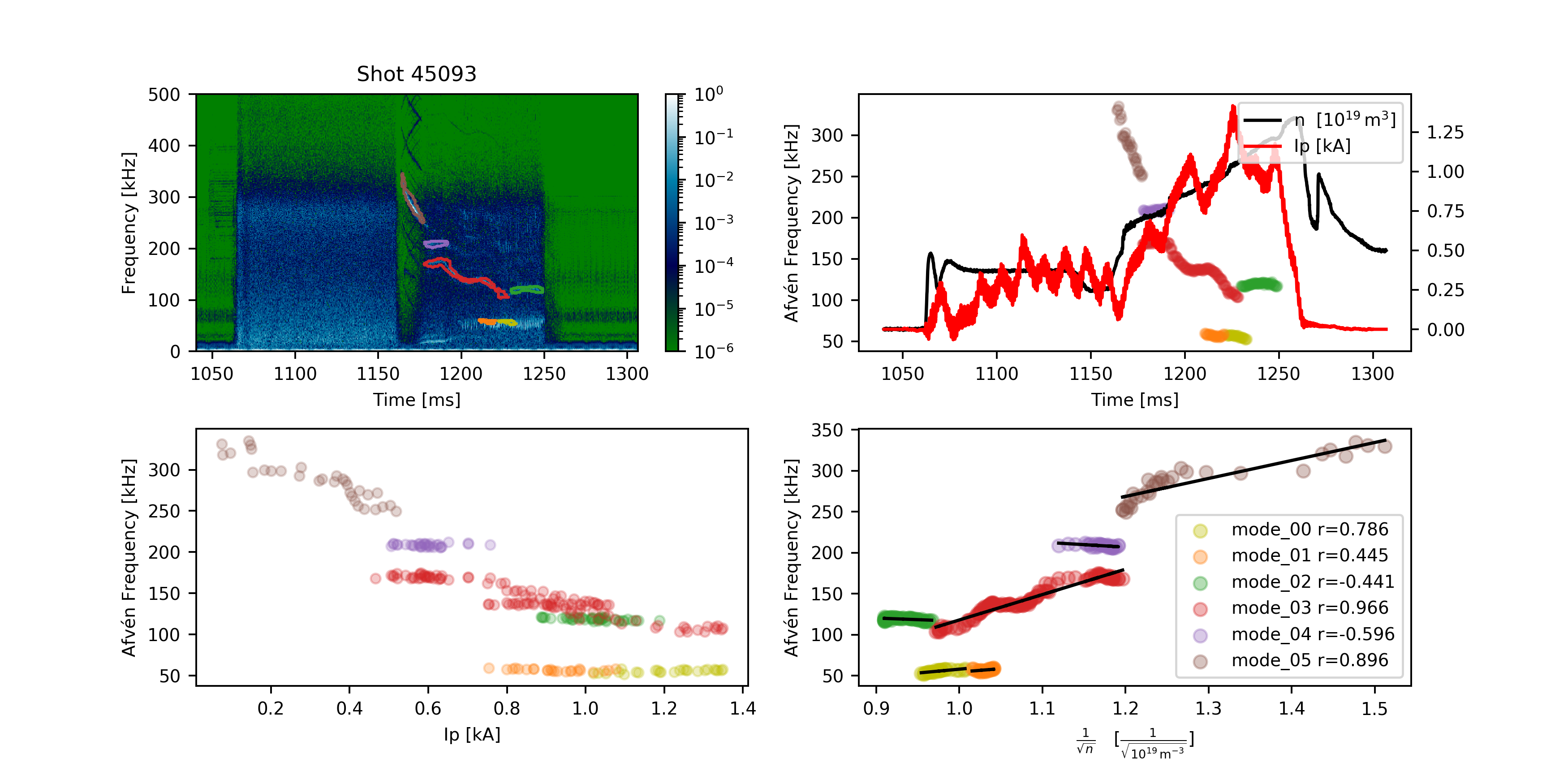}
  \caption{Examples of Alfv\'en modes detected with the image segmentator after some post processing. Since the plasma current is small and evolves slowly, the $f_{\mathrm{Alfven}} \sim n^{-1/2}$ condition is nicely fulfilled.}
  \label{fig_alfven_examples}
\end{figure}

Once we had a functional mode detector, we applied it to all discharges we have to show its features and possibilities. First we show some examples of the detection of Alfv\'en modes using the correlation coefficient. In Fig. \ref{fig_alfven_examples} we can see three discharges with modes with correlation coefficient $\gtrapprox 0.9$. There are four plots of every TJ-II shot. Starting on the upper left corner and following clockwise, we plot the spectrogram with the identified modes; density and current time traces, and the evolution of the mode frequencies; scatter plot of the mode frequency versus $n^{-1/2}$ and the corresponding linear correlation coefficient; and finally a scatter plot of the mode frequency and the plasma current.

\begin{figure}[ht]
  \centering
  \includegraphics[width=12cm]{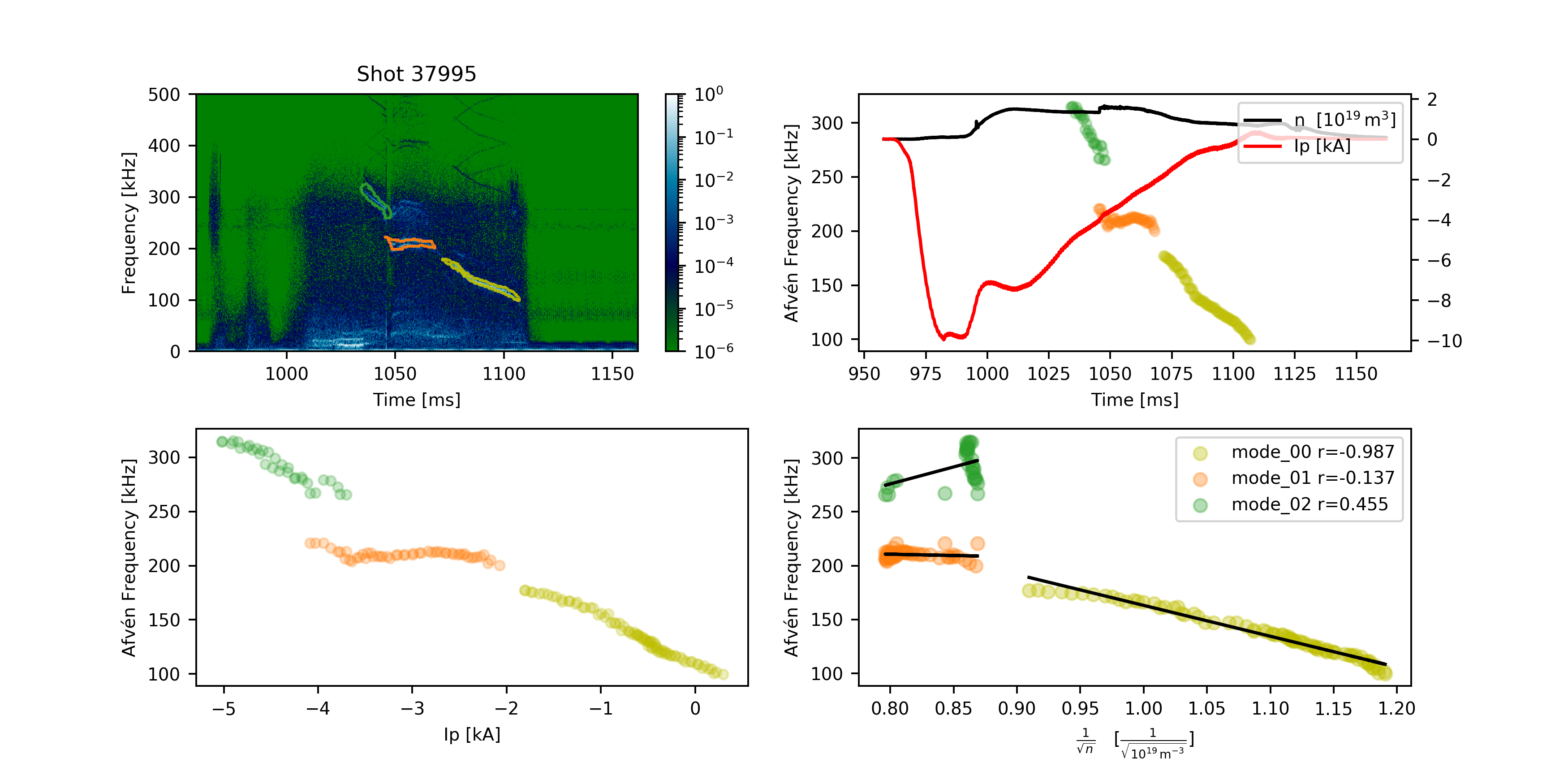}
    \includegraphics[width=12cm]{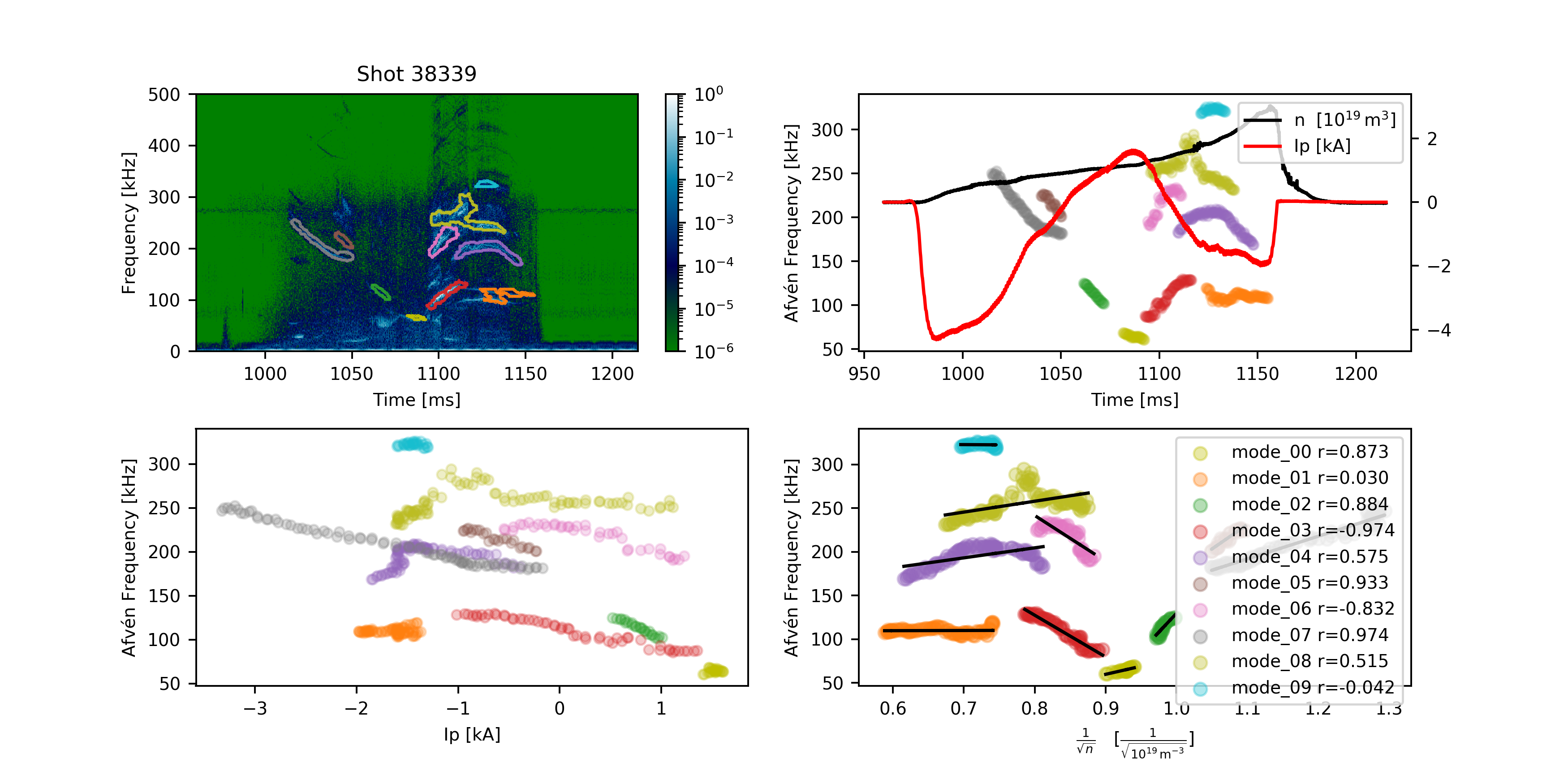}
  \includegraphics[width=12cm]{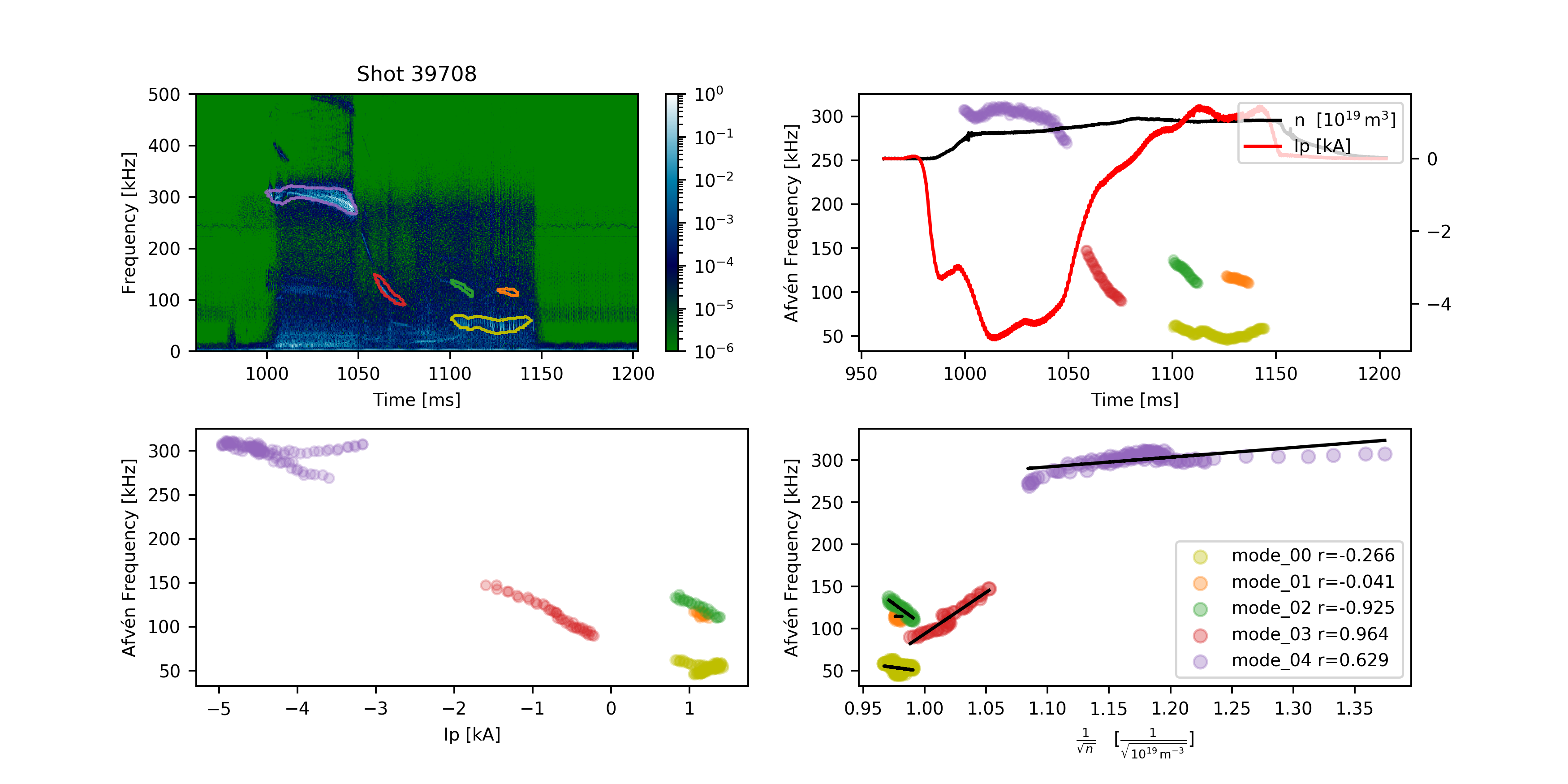}
  \caption{Examples in which, in spite the modes living in the Alfvenic frequency range, the $f_{\mathrm{Alfven}} \sim n^{-1/2}$  condition is not fulfilled without taking into account the strong evolution of the plasma current along the discharge.}
  \label{fig_NOalfven_examples}
\end{figure}

Although in most detected modes the mentioned correlation coefficient is high, we can also find some cases with MHD activity with frequency over 50 kHz in which the plasma current changes abruptly and modifies the Shear Alfv\'en Spectrum due to its impact on the rotational transform. In those cases the direct comparison of frequency with $n^{-1/2}$ is misleading because the plasma current variation must be taken into account \cite{Sun_NF_2015}. Some examples of this type of modes are shown in Fig. \ref{fig_NOalfven_examples}.

All mode information is stored in a non-relational database (MongoDB) for further queries and analysis. In this way, any user can filter out the modes according to his criteria and interests. For example, we can obtain the distribution of several numerical features of the modes that were not previously calculated in TJ-II.

\begin{figure}[ht] 
\subfloat[]{\includegraphics[width = 7cm]{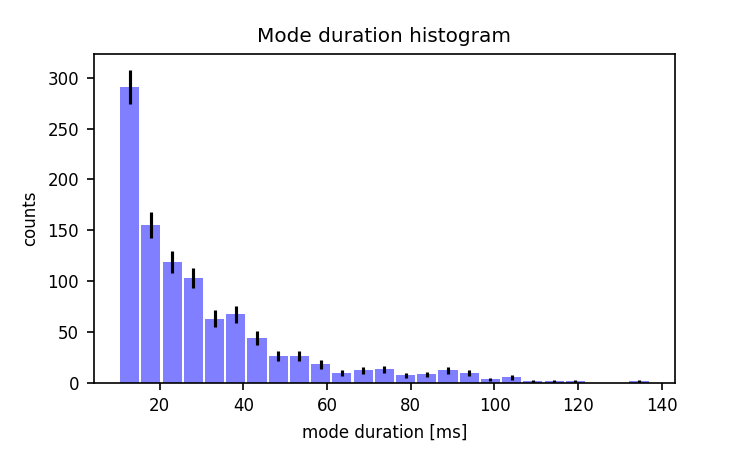}}
\subfloat[]{\includegraphics[width = 7cm]{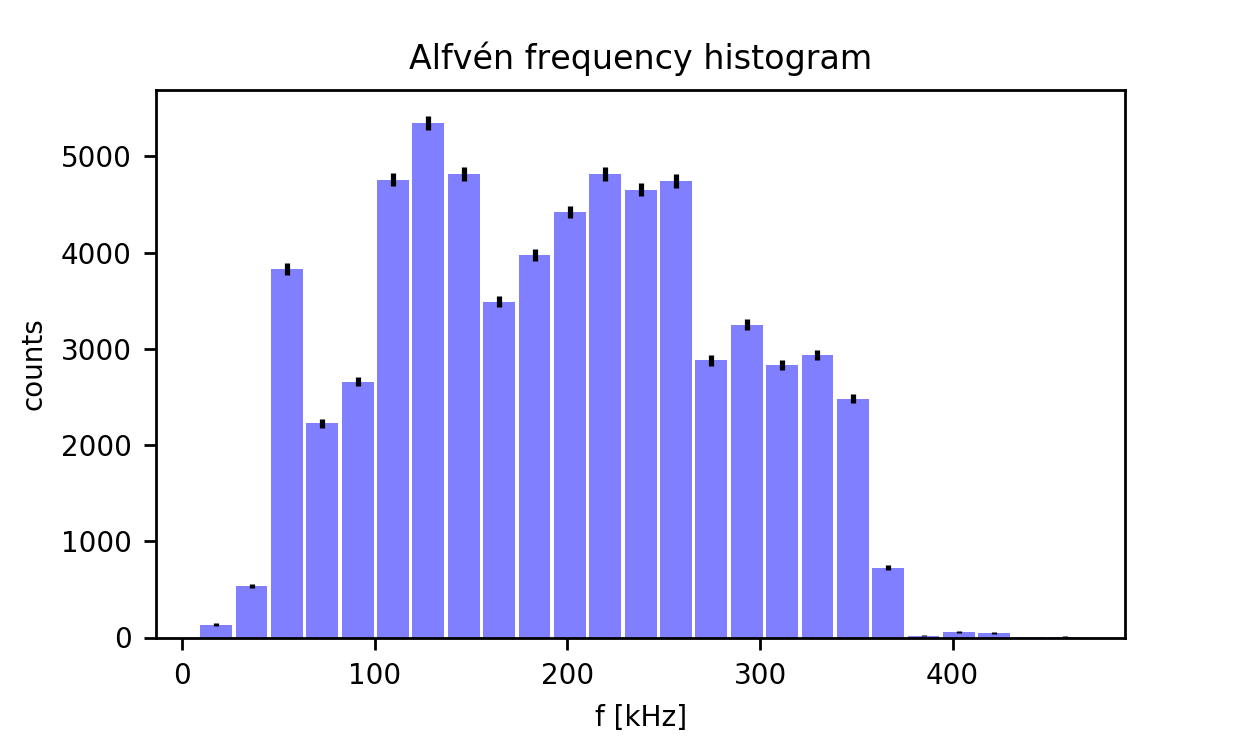}}\\
\subfloat[]{\includegraphics[width = 7cm]{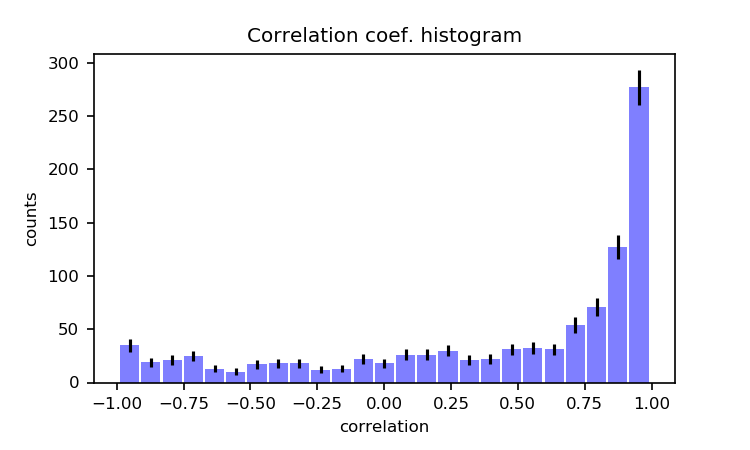}}
\subfloat[]{\includegraphics[width = 7cm]{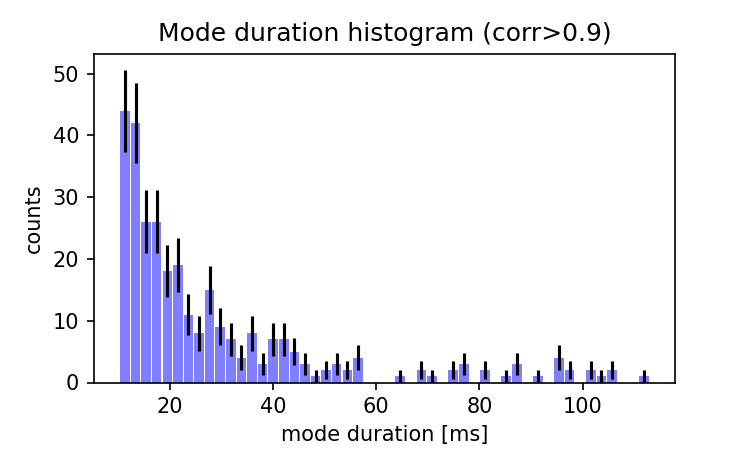}}\caption{Examples of data mining results: a) Distribution of the mode duration in the dataset b) Distribution of the mode frequencies in the dataset c)Distribution of the correlation coefficient between the mode frequency and $n^{-1/2}$ and d) Distribution of the mode duration restricted to modes with correlation coefficient between the frequency and $n^{-1/2}$ larger than 0.9.}
\label{fig_data_mining}
\end{figure}

In Fig. \ref{fig_data_mining} we plot some data mining results obtained from the whole dataset of detected modes. We can see a histogram of the duration of the 989 detected modes (a). Most modes last for a few tens of milliseconds, with some of them lasting more than 100 ms (recall that we neglect any mode shorter than 10 ms). A similar histogram (b) showing all detected frequencies. Most modes concentrate in frequencies between 100 and 350 kHz, with peaks around 130 and 250 kHz, are shown. In (c), the histograms of the correlation coefficient between the frequency and $n^{-1/2}$ for each mode. This is very peaked around 1, meaning that most of the modes detected have Alfv\'en features and relatively small and slow-evolving plasma current. In (d) we plot the distribution of the mode duration restricted to modes with correlation coefficient larger than 0.9. Those examples show the potential of our method to extract information from a big dataset of spectrograms and identify experimental features and/or systematic behavior of the MHD oscillations.

\subsection{Detector sensitivity}
The sensitivity of the mode detector can be adjusted for future applications. Recall that, given an input spectrogram, our binary segmentator estimates the probability of each pixel to belong to a mode. The default value of the acceptance probability is 0.5, but we can reduce it and increase the sensitivity (and thus increase the number of false positives). If we increase this threshold we change the behavior and operate the segmentator in a more conservative way, rising the number of false negatives. A visual inspection of some examples of the dataset can help to adjust this value depending on the interests of the user.

\begin{figure}[ht]
  \centering
  \includegraphics[width=12cm]{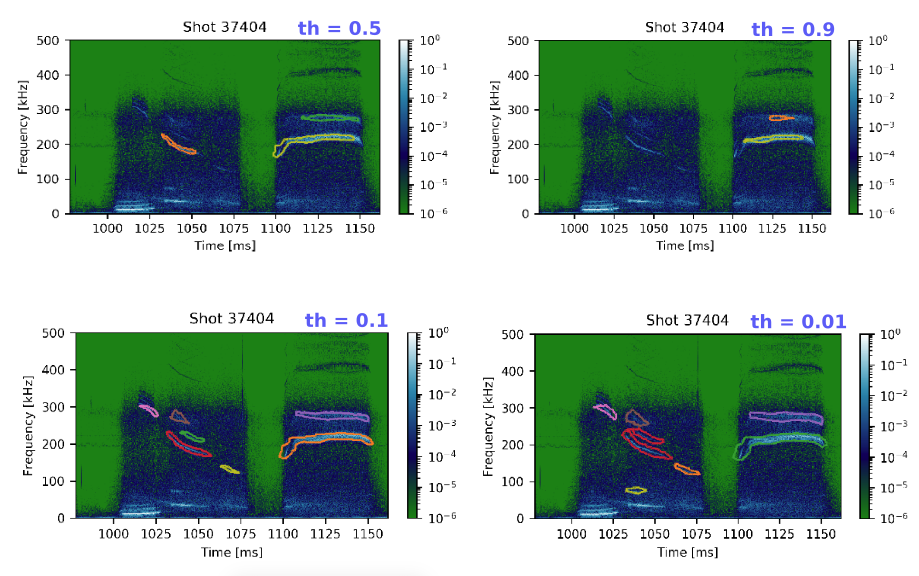}
  \caption{With thresholds larger than 0.5 we become more strict and conservative, reducing the number of false positives but missing existing modes. When this threshold is smaller, we increase the sensitivity of the detector at the prize of, possibly, increasing the number of false positives.}
  \label{fig_sensitivity}
\end{figure}

In Fig. \ref{fig_sensitivity} we show four predictions of the same spectrogram using four different sensitivity thresholds. We can see that as we decrease the threshold, the area identified as mode increases and vice-versa. It is important to notice that there is no need to train again the neural network to use this functionality.

\section{Conclusions}
\label{sec_conclusions}
In this work we have developed an analysis tool based on Deep Learning to identify oscillation modes in Mirnov coil spectrograms measured in confined plasma discharges. The target is to find series of frequency and time pairs that locate oscillations modes in a spectrogram.  We have trained such tool with a manually annotated dataset of 500 spectrograms from the TJ-II stellarator database. The detector turns out to work well, giving AUC scores larger than 0.99 (excellent) and IoU larger than 0.4 (enough for our purposes). To the authors' knowledge, this is the first time that such techniques are applied to this kind of dataset, and it has turned out to be a valid tool that can save time to the experimental plasma physicists.

We can use it to effectively mine any set of discharges in TJ-II, creating a database of modes with their physical characteristics. Once the segmentator is trained, the database creation takes $\approx 7$ seconds per spectrogram on a desktop computer. Then this information can be used to identify patterns, physical behavior or check hypothesis.

It is worth to mention that the software developed and methods employed can be applied to other facilities and diagnostics that produce images and/or spectrograms like bolometry, Langmuir probes, heavy ion beam probe, etc. Transfer learning techniques can be used here to speed up new trainings.

\begin{figure}[ht]
  \centering
  \includegraphics[width=12cm]{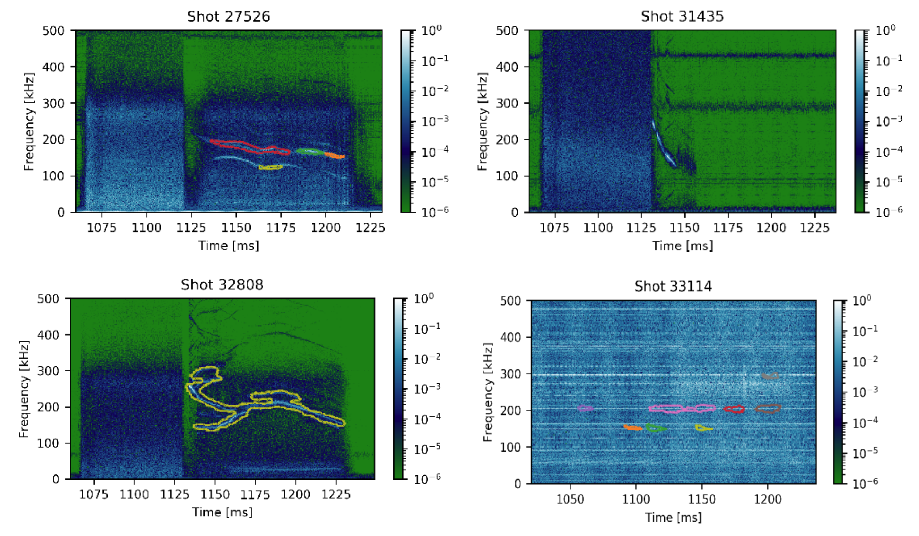}
  \caption{Some errors that sometimes appear in the mode detection. Starting from the upper-left picture and following clockwise, we can see that the segmentator missed parts of a long mode, missed a whole mode, merged many modes in a very complex shape and detected ghost modes in a failed discharge.}
  \label{fig_detector_errors}
\end{figure}

For the sake of completeness we also give some of the few examples where the mode detector does not work properly. We observe in Fig. \ref{fig_detector_errors} modes that are misdetected and modes that are merged by the detector and identified as a single mode. Some of the mentioned errors might be avoided simply by increasing the dataset size and train the neural network again. Usually, the neural networks increase their performance with the dataset size, as opposed to other simpler machine learning models that saturate after certain dataset size. The problem with modes that crosses each other can be solved changing the neural network architecture so the segmentator is able to resolve the instance map, i.e. identify distinct modes. But this requires new annotation and training processes and it is left for future work.

Finally we can say that, overall, the mode detector gives reasonable good results and it is a potential tool for automatic analysis of spectrograms.

\section*{Acknowledgments}
This work was partially funded by the Spanish Ministry of Science and Innovation under contract number FIS2017-88892-P and RTI2018-096006-B-I00(CODEC-OSE), and  by the  Comunidad  de  Madrid  CABAHLA  project  (S2018/TCS-4423). Simulations carried out  made  also  use  of  the  computing  facilities  provided  by  the  CYTED  Network RICAP (517RT0529). 

\bibliography{references}{}

\begin{thebibliography}{10}

\bibitem{Helander_book}
P. Helander and D.~J. Sigmar, {\em Collisional Transport in Magnetized Plasmas}
  (Cambridge University Press, Cambridge, 2001).

\bibitem{Balescu_book}
R. Balescu, {\em Transport Processes in Plasmas: Neoclassical Transport Theory}
  (Elsevier Science Ltd, Netherlands, 1988).

\bibitem{Vlad_1999}
G.Vlad, F.Zonca, and S.Briguglio, Rivista del Nuovo Cimento {\bf 22},
  (1999).

\bibitem{Mirnov_Soviet_1971}
S. Mirnov and I. Semenov, Soviet Physics JETP {\bf 33, 6},    (1971).

\bibitem{Otsu_1979}
N. Otsu, IEEE Transactions on Systems, Man and Cybernetics {\bf 9},  62
  (1979).

\bibitem{Goodfellow_book}
I. Goodfellow, Y. Bengio, and A. Courville, {\em Deep Learning} (MIT Press,
  USA, 2016), \url{http://www.deeplearningbook.org}.

\bibitem{Minaee2020ImageSU}
S. Minaee {\it et~al.}, ArXiv {\bf abs/2001.05566},    (2020).

\bibitem{Simonyan2015VeryDC}
K. Simonyan and A. Zisserman, CoRR {\bf abs/1409.1556},    (2015).

\bibitem{howard2017mobilenets}
A.~G. Howard {\it et~al.},   (2017), cite arxiv:1704.04861.

\bibitem{repo_segmentator}
D. Gupta and R.~J. Wala, Image Segmentation Keras : Implementation of Segnet,
  FCN, UNet, PSPNet and other models in Keras.,
  https://github.com/divamgupta/image-segmentation-keras, 2020.

\bibitem{Jimenez-Gomez_FST_2007}
R. Jimenez-Gomez {\it et~al.}, Fusion Sci. Technol. {\bf 51},  20  (2007).

\bibitem{Jimenez-Gomez_NF_2011}
R. Jim{\'{e}}nez-G{\'{o}}mez {\it et~al.}, Nuclear Fusion {\bf 51},  033001
  (2011).

\bibitem{BRADLEY19971145}
A.~P. Bradley, Pattern Recognition {\bf 30},  1145   (1997).

\bibitem{kingma2014method}
D.~P. Kingma and J. Ba, 3rd International Conference for Learning
  Representations  (2015), cite arxiv:1412.6980.

\bibitem{Sun_NF_2015}
B. Sun, M. Ochando, and D. L{\'{o}}pez-Bruna, Nuclear Fusion {\bf 55},  093023
  (2015).

\end{thebibliography}
\bibliographystyle{prsty}
\end{document}